\documentclass{article}
\usepackage{graphicx}  
\usepackage{amsmath}   
\usepackage{amssymb}   
\usepackage[active]{srcltx}
\usepackage{subfig}
\def\eq#1{{Eq.~(\ref{#1})}}

\def \lp {\ell_0}
\def\DM{\mathrm{d}}
\title{Spacetime with zero point length is two-dimensional at the Planck scale}
\author{T. Padmanabhan
\footnote{paddy@iucaa.in} 
\hskip 0.05 truecm,
Sumanta Chakraborty
\footnote{sumanta@iucaa.in;~~sumantac.physics@gmail.com} 
\\
{\small{\it IUCAA, Post Bag 4, Ganeshkhind,}}
{\small{\it Pune University Campus, Pune 411 007, India}}\\
and
\\
Dawood Kothawala
\footnote{dawood@physics.iitm.ac.in}
\\
{\small{\it Department of Physics, IIT Madras, Chennai-600 036, India}}
}
\date{}  
\begin{document}
  
\maketitle
\begin{abstract}
\noindent It is generally believed that any quantum theory of gravity should have a generic feature --- a quantum of length. We provide a physical ansatz to obtain an effective \emph{non-local} metric tensor starting from the standard metric tensor such that the spacetime acquires a zero-point-length $\lp$ of the order of the Planck length $L_{P}$. This prescription leads to several remarkable consequences. In particular, the Euclidean volume $V_D(\ell,\lp)$ in a $D$-dimensional spacetime of a region of size $\ell $ scales as $V_D(\ell, \lp) \propto \lp^{D-2} \ell^2$ when $\ell  \sim \lp$, while it reduces to the standard result $V_D(\ell,\lp) \propto \ell^D$ at large scales ($\ell  \gg \lp$). The appropriately defined effective dimension, $D_{\rm eff} $, decreases continuously from $D_{\rm eff}=D$ (at $\ell  \gg \lp$) to $D_{\rm eff}=2$ (at $\ell \sim \lp$). This suggests that the physical spacetime becomes essentially 2-dimensional near Planck scale.
\end{abstract}
The existence of a fundamental length scale, which provides an ultimate lower bound to measurement of spacetime intervals, is a model-independent feature of quantum gravity, since various approaches to quantum gravity lead to this result. The appearance of such a minimum length scale actually derives from basic principles of quantum mechanics, special and general relativity, and hence is often considered as a robust feature of quantum gravity. This result can have important consequences, e.g., notions of causality or distance between two events cannot be expected to have a continuous behavior at this length scale \cite{zero}. 

It seems, therefore, natural to look for a description of spacetime at the mesoscopic scales which interpolate between Planck scale and low energy scales, by incorporating the effect of introducing a zero point length to the spacetime through modification of the spacetime metric $g_{ab}$ to some suitable object $q_{ab}$, which we shall refer to as the \textit{qmetric}. The $q_{ab}$ must be constructed such that the geodesic distance between two points computed using $q_{ab}$ acquires a lower bound \cite{tpdk1}.  If we can determine the qmetric in terms of $g_{ab}$, then we can compute all other geometrical variables (like, for e.g., curvature tensor) by using the qmetric in the place of $g_{ab}$ in the relevant expressions. Such a procedure is necessarily approximate --- compared to a fully rigorous non-perturbative quantum gravitational approach --- but will surely capture some of the effects at the mesoscopic scales which interpolate between Planck scale and low energy scales at which the 
classical metric provides an adequate description. 

Unfortunately, we cannot use perturbation techniques to compute $q_{ab}$ for a given classical geometry described by a $g_{ab}$. In fact, we would expect it to be non-local and singular at any given event. In the absence of explicit computability, we need a physically motivated ansatz to relate  $q_{ab}$ to $g_{ab}$. Such an ansatz was introduced and described in fair detail in \cite{tpdk1}. The essential idea was to recognize that a primary effect of quantum gravity will be to endow spacetime with a zero-point length \cite{zero} by modifying the geodesic interval $\sigma^2(x,x')$ between two events $x$ and $x'$ in a Euclidean spacetime  to a form like $\sigma^2 \to \sigma^2 + \lp^2$. More generally, we will assume that a key effect of quantum gravity is to modify $\sigma^2 \to {\mathcal F}(\sigma^2)$  where the function ${\mathcal F}(\sigma^2)$ satisfies 
the constraint ${\mathcal F}(0) = \lp^2$. While most of our results are insensitive to the explicit functional form of ${\mathcal F}(\sigma^2)$, for illustration, we will use ${\mathcal F}(\sigma^2) = \sigma^2 + \lp^2$. 

This approach essentially treats the  $\sigma^2(x,x')$ as more fundamental than the metric $g_{ab}(x)$ itself. As we have already emphasized there is  considerable amount of evidence for the existence of such a zero-point-length of spacetime; there have also been arguments in the literature suggesting that $g_{ab}$ may not be the most primitive variable to use while studying quantum gravitational effects (see e.g., \cite{thooft}). The key claim being, while we may not know how quantum gravity modifies the classical metric a priori, we \textit{do} have an indirect handle on it, if we assume that quantum gravity introduces a zero-point length to the spacetime in the manner described above.

One can determine \cite{dk0,dk2} the form of $q_{ab}$ for a given $g_{ab}$ by using the requirements that 
(i) it should lead to a spacetime interval with a zero-point-length and 
(ii) the two-point function describing small perturbations of the metric should have a regularized non-singular coincidence limit. It was shown in the earlier works \cite{dk0,dk2} that these conditions allow us to determine $q_{ab}$ uniquely in terms of $g_{ab}$ (and its associated geodesic interval $\sigma^2$). We find that 
\begin{align}
q_{ab}=Ah_{ab}+ B n_{a}n_{b};\qquad q^{ab}=\frac{1}{A}h^{ab}+\frac{1}{B}n^{a}n^{b}
\end{align}
where $h_{ab}=g_{ab}-n_a n_b$, and
\begin{align}
B=\frac{\sigma ^{2}}{\sigma ^{2}+\lp^{2}};\qquad A=\left(\frac{\Delta_{\phantom{F}}}{\Delta _{\mathcal F}}\right)^{2/D_{1}}\frac{\sigma ^{2}+\lp^{2}}{\sigma ^{2}};\qquad n_a=\dfrac{\nabla _{a}\sigma ^{2}}{2\sqrt{\sigma ^{2}}}
\end{align}
with $\Delta$ being the Van Vleck determinant related to the geodesic interval $\sigma^2 $ by
\begin{align}
\Delta (x,x')=\frac{1}{\sqrt{g(x)g(x')}}\textrm{det}\left\lbrace \frac{1}{2} \nabla _{a}^{x}\nabla _{b}^{x'}\sigma ^{2}(x,x') \right\rbrace
\end{align}
$\Delta_\mathcal F$ is defined by replacing $\sigma^{2}$ with $\mathcal{F}(\sigma^{2})$ in the series expansion of $\Delta$ (see \cite{dk2}). We interpret $q_{ab}$ as the effective spacetime metric incorporating some of the non-perturbative effects of quantum gravity at Planck scales. As described in detail in the previous works \cite{tpdk1,tpdk2} the qmetric has the following properties:

(1) Unlike $g_{ab}(x)$, the qmetric $q_{ab}(x,x')$ is a bi-tensor depending on two events $x,x'$ through $\sigma^2$. It is easy to show that this non-locality is essential if spacetime has to acquire a zero-point length. Almost by definition, any local metric will lead to a geodesic interval which vanishes in the limit of $x\to x'$.

(2) $q_{ab}$ reduces to the background metric $g_{ab}$ in the limit of $\lp^2 \to 0$. That is, in the classical limit of $\hbar \to 0$, qmetric reduces to the standard metric as we would expect. 

(3) In the opposite limit of $(\sigma^2 / \lp^2) \to 0$ the qmetric is singular at all events. This is natural when we interpret qmetric as the metric of the mesoscopic spacetime; we would not expect it to be well defined at any given event and will require some kind of smearing over Planck scales for it to be meaningful.

(4) If the background metric is flat, then the qmetric is also flat, i.e., there exists a coordinate mapping from $q_{ab} \to \eta_{ab}$. This is, however, rather subtle because the coordinate transformation is, in fact, singular in the coincidence limit in (regular) Cartesian coordinates, and the mapping effectively removes a geodesic region of size $\lp$ from the spacetime around all events. 

(5) Let $\Phi[g_{ab}(x)]$ be a scalar  constructed from the background metric and possibly several of its derivatives, like, for example, the Ricci scalar $R[g_{ab}(x)]$. 
We can now compute the corresponding (bi)scalar $\Phi[q_{ab}(x,x');\lp^2]$ for the qmetric by replacing $g_{ab}$ by $q_{ab}$ in $\Phi[g_{ab}(x)]$ and evaluating all derivatives at $x$ keeping $x'$ (``base point'') fixed. We interpret the value of this scalar by taking the limit $x\to x'$ in this expression keeping $\lp^2$ non-zero. As noted in \cite{tpdk1,tpdk2}, several useful scalars like $R$, $K$ etc. remain finite and local in this limit even though the qmetric itself is singular when $x\to x'$ with non-zero $\lp^2$. This arises from the algebraic fact that the following two limits do not commute:
\begin{equation}
\lim_{\lp^2\to 0}\, \lim_{x\to x'} \Phi[q_{ab}(x,x');\lp^2]\neq  \lim_{x\to x'}\,\lim_{\lp^2\to 0} \Phi[q_{ab}(x,x');\lp^2]
\label{noncommute}
\end{equation} 

All these computations are most easily performed \cite{tpsc1} by choosing a synchronous coordinate system for the background metric which can always be done in a local region. In this coordinate system, the equi-geodesic surface $\sigma = $ constant, which is at a constant geodesic distance from the base point, has a simple description. The expressions for a few geometrical variables, computed by this technique with the limit $x\to x'$ taken in the end, are given in the appendix. (Some of these results have been independently obtained by a more detailed computation, without using the synchronous frame, in \cite{dk2}.)
These facts show that we have a well-defined procedure for doing the computations in the mesoscopic spacetimes using the qmetric.  More details regarding this approach will be presented elsewhere \cite{tpsc1}. 

In this work, we will  concentrate on one key effect of such a renormalization of the spacetime metric. This relates to the volume of the geodesic ball:  
\begin{equation}
V_D(\ell,\lp)\equiv  \int_{\sigma \leq \ell } \DM \sigma \DM \Omega _{D-1}\, \sqrt{q}         
\end{equation} 
enclosed by an equi-geodesic surface of size $\sigma = \ell $ in the $D$-dimensional Euclidean spacetime. Here, $q$ is the determinant of the qmetric with 
\begin{align}
\sqrt{q} = \left(\frac{\Delta_{\phantom{F}}}{\Delta _{\mathcal{F}}}\right)\left(\frac{\sigma ^{2}+\lp^{2}}{\sigma ^{2}} \right)^{D_{2}/2}\sqrt{g};
\quad D_2\equiv (D-2)
\end{align}
Classically we would expect the scaling $V_D(\ell )\propto \ell^D$ for sufficiently small $\ell $ when we can ignore the scales involved in spacetime curvature. Let us now evaluate the same quantity for the mesoscopic spacetime. Using the synchronous coordinates, a straightforward (though lengthy) calculation, gives the following result: 
\begin{equation}
\sqrt{q}= \sigma \left(\sigma ^{2}+\lp^{2}\right)^{D_{2}/2}\left(1-\frac{1}{6}\mathcal{S}(x')\left(\sigma ^{2}+\lp^{2}\right) + \cdots \right)
\end{equation}
where $\mathcal{S}\equiv R_{ab}n^{a}n^{b}$ is a scalar constructed from the background $g_{ab}$ and is evaluated at base point $x'$. (Though we only need the leading order term, we have displayed next order term in the expansion which contribute in the limit of $\lp^2/\sigma^2 \to 0$.) The $n_{a}$ is the tangent to the geodesic and  hence depends on the geodesic that connects the base point $x'$ to the point $x$ on the equi-geodesic surface. Thus integrating over the angular coordinates on the equi-geodesic surface amounts to averaging  $n_{a}$ over the solid angle. [Such an the angular integral over $n^{a}n^{b}$ leads to, $(\Omega /D)g^{ab}$, where $\Omega=2\pi^{D/2}\Gamma (D/2)^{-1}$ (see e.g.,\cite{Gray1973})]. A straightforward integration now gives the result, to the same order of accuracy. as:
\begin{align}
\label{onesix}
V_D(\ell,\lp) =\frac{\Omega}{D}\left\lbrace \left(\ell^{2}+\lp^{2}\right)^{D/2}-\lp^{D}\right\rbrace
-\frac{\Omega}{6D(D+2)}R(x')\left\lbrace \left(\ell^{2}+\lp^{2}\right)^{(D+2)/2}-\lp^{D+2}\right\rbrace 
\end{align}
where  and $R$ is the Ricci scalar for the bare metric. 

Let us now consider the two relevant limits of this expression. First, when $(\lp/\ell ) \to 0$, we get
\begin{align}
\lim _{\lp\rightarrow 0} V_D(\ell,\lp) =\frac{\Omega }{D}\ell^{D}\left[1-\frac{1}{6(D+2)}R(x')\ell^{2}\right]
\end{align}
which shows that the volume scales as $\ell^D$ except for curvature induced corrections captured by the second term. This correction is a standard result known in differential geometry \cite{Gray1973}.  What is more interesting is the limit of $(\ell /\lp) \to 0$, when we get
\begin{align}
\lim _{\ell  \rightarrow 0} V_D(\ell,\lp)=\frac{\Omega }{2}\lp^{D}\left(\frac{\ell^{2}}{\lp^{2}} \right)\left[1-\frac{1}{6D}R(x')\lp^{2}\right] \to \frac{\Omega }{2}\lp^{D}\left(\frac{\ell^{2}}{\lp^{2}} \right) \propto \ell^2
\end{align}
This suggests the remarkable possibility that the existence of zero-point-length makes the physical spacetime essentially  2-dimensional near Planck scale. A convenient measure of such a dimensional reduction is provided by the quantity\footnote{It might seem that a simpler definition is just $D_{\rm eff} = ({\DM \ln{ V_D(\ell,\lp) } }/{\DM \ln \ell })$. However, in any $D$-dimensional curved space with a smooth metric $g_{ab}$, this definition will  give $D$ only for $R\ell^2\ll 1$; that is at scales small compared to curvature scale.  The role of $V_D(\ell,\lp =0)$ in the definition is to  remove the contribution from the background curvature to $D_{\rm eff}$ thereby ensuring that $D_{\rm eff}=D$ when $\ell_0=0$. So any deviation of $D$ from $D_{\rm eff}$ arises only due to the existence of the zero-point-length.}
\begin{align}
D_{\rm eff} = D + \dfrac{\DM}{\DM \ln \ell }\left\lbrace \ln \left(\frac{ V_D(\ell,\lp)}{V_D(\ell,\lp=0)}\right) \right\rbrace
\end{align}
which, using the above expressions, decreases from $D_{\rm eff}=D$ for large $\ell$ to $D_{\rm eff}=2$ as $\ell \to 0$. 

There have been several indications from various approaches to quantum gravity that spacetime might ``look" two dimensional when probed at extremely small scales (most of these arguments refer to the so called {\it spectral dimension} defined via the process of random walk). Carlip \cite{carlip} has given several independent set of arguments suggesting that such a dimensional reduction may be an inevitable feature of quantum gravity. Somewhat closer in spirit to the ideas presented here seems to be the results of \cite{cdt}, where the 
authors argue that dimensional reduction to $D=2$ might provide a mechanism by which quantum gravity ``self-renormalizes" at Planck scale. A geometrically similar approach (but based on quantization of area) in the context of loop quantum gravity and spin-foam models, was discussed in \cite{modesto}, while similar result was obtained in the context of generalized uncertainty principle in \cite{Viqar}. The current study links such a dimensional reduction to the existence of zero-point-length in spacetime, \textit{independent of any specific model of quantum gravity}.

To understand the \textit{algebraic} origin of this result, one could study the case of a flat spacetime in synchronous coordinates in which $\sigma$ denoting the radial geodesic distance from a chosen origin.  In this case, the qmetric leads to the line element of the form (when $D=4$)
\begin{align}
\DM s^{2}=\frac{1}{A(\sigma)} \DM \sigma ^{2}+A(\sigma)\sigma ^{2} \DM\Omega _{3}^{2} ; \quad A = 1+ \frac{\lp^2}{\sigma^2}
\end{align}
which transforms to ${ds}^{2}=d\bar{\sigma}^{2}+\bar{\sigma}^{2}d\Omega _{3}^{2}$
with $\bar{\sigma}=\sqrt{\sigma ^{2}+\lp^{2}}$. In regular Cartesian coordinates, this transformation is {\it singular} in the coincidence limit (see Appendix A of \cite{dk0}) and is equivalent to removing a ``hole'' of radius $\lp$ from the manifold in a specific manner. In fact, in flat spacetime, we will get the result $V_D(\ell,\lp) = (\Omega/D)[\left(\ell^{2}+\lp^{2}\right)^{D/2}-\lp^{D}]$ which clearly has the limits $V_D(\ell,\lp) \propto \ell^D$ (when $(\lp/\ell ) \to 0$) and $V_D(\ell,\lp) \propto \ell^2$ (when $(\ell /\lp) \to 0$). This clarifies the algebraic origin of the result.

We believe the conjecture, that the mesoscopic spacetime is described by the qmetric, is a powerful and useful one. It is well motivated by the existence of the zero-point-length in the spacetime and leads to well defined computation rules which can incorporate the effects of quantum gravity at mesoscopic scales without us leaving the comfort of a continuum differential geometry (albeit one involving a singular and non-local metric). Its power is seen once again in the current work, where it leads to a definite conclusion that the effective  dimensionality of the spacetime at Planck scales is $D_{\rm eff}=2$. This opens up useful further avenues of exploration.
\paragraph*{Acknowledgements}
Research of T.P is partially supported by J.C. Bose research grant of DST, Govt. of India. Research of S.C is funded by SPM Fellowship from CSIR, Govt. of India. We thank Krishnamohan Parattu, Suprit Singh and Kinjalk Lochan for helpful discussions. 
\paragraph*{Appendix}
In this appendix we will briefly illustrate how the prescription of qmetric works by describing the key steps in computing $R$ and $K$ for the qmetric; the general derivation and expressions can be found in \cite{dk2}. Here, we will sketch an easier method based on using the synchronous frame for the background metric (the details will be given in \cite{tpsc1}). The qmetric line element in the synchronous frame turns out to be,
\begin{align}
\DM s^{2}=\frac{1}{A(\sigma)}\left[\DM \sigma ^{2}+A^{2}(\sigma)\left(\frac{\Delta_{\phantom{F}}}{\Delta _{\mathcal F}}\right)^{2/D_{1}}h_{\alpha \beta} \DM x^{\alpha} \DM x^{\beta}\right];\qquad A=\frac{\sigma ^{2}+\lp^{2}}{\sigma ^{2}}
\end{align}
where $D_{1}=D-1$. Then  the Ricci biscalar $R[q_{ab}]$ can be computed using the usual formula which relates the metric to Ricci scalar with all derivatives acting on $x$. We then get:
\begin{align}
R_{q}&=\frac{1}{A}R_{\Sigma} \left(\frac{\Delta_{\phantom{F}}}{\Delta _{\mathcal F}}\right)^{-2/D_1} - \frac{D_1D_2}{\mathcal F(\sigma ^{2})}+4(D+1)\dfrac{d\ln \Delta _{\mathcal F}}{\DM \mathcal F}
\nonumber
\\
&-\frac{\mathcal F}{\sigma ^{2}}\left(K_{ab}K^{ab}-\frac{1}{D_{1}}K^{2}\right)
+4 \mathcal F \left[-\frac{D}{D_{1}}\left(\frac{d\ln \Delta _{\mathcal F}}{\DM \mathcal F}\right) ^{2}+2\dfrac{\DM^{2}\ln \Delta _{\mathcal F}}{\DM \mathcal F^{2}}\right]
\label{r1}
\end{align}
where again $D_{2}$ stands for $D-2$ and $R_{\Sigma}$ stands for the curvature scalar on $\sigma =\textrm{constant}$ surface. In the final expression we interpret $\sigma(x,x')$ as the geodesic distance in the background metric. Then using expressions for derivatives of the van Vleck determinant, we first take the coincidence limit $\sigma ^{2}\rightarrow 0$
keeping $\lp$ finite to obtain (with $\mathcal{S}\equiv R_{ab}n^{a}n^{b}$):
\begin{equation}
\lim _{\sigma \rightarrow 0}R_{q}=D\mathcal{S}+\frac{2}{3}(D+1)\left(n^{a}\nabla _{a}\mathcal{S}\right)\lp+\mathcal{O}(\lp^{2})
\end{equation} 
If we now take $\lp\rightarrow 0$ we get,
\begin{align}
\lim _{\lp\rightarrow 0}\lim _{\sigma \rightarrow 0}R_{q}=D\mathcal{S}
\end{align}
where all the objects on the right hand side is being determined by the bare metric $g_{ab}$ (see also \cite{dk2}). Note that if we first take the limit $\lp\rightarrow 0$ in \eq{r1}, we would obtain the Ricci scalar for the bare metric and the second limit of $\sigma^2\to0$ becomes vacuous. Hence the two limits (a) $\sigma \rightarrow 0$ and (b) $\lp\rightarrow 0$  do not commute, as noted in  Eq. (\ref{noncommute}). 

Similar result holds for trace of extrinsic curvature $K$ of the equi-geodesic surface evaluated for the qmetric. There we readily obtain leading terms in orders of $\lp^{2}$ to be,
\begin{align}
\lim _{\sigma \rightarrow 0} K_{q}=\frac{D_{1}}{\lp}-\frac{\lp}{3}\mathcal{S}
\label{k1}
\end{align}
In which $\mathcal{S}=R_{ab}n^{a}n^{b}$, evaluated for the bare metric, appears again.  This quantity $\mathcal{S}$ is the entropy density of the null surfaces used in the emergent gravity paradigm \cite{emgrav}. The first term in \eq{k1} is a zero point entropy density of the spacetime and is closely related to the possible solution to the cosmological constant problem \cite{tpdk2,tphp}.


 \end{document}